\documentclass{wscpaperproc}
\usepackage{latexsym}
\usepackage{graphicx}
\usepackage{mathptmx}
\usepackage[T1]{fontenc}

\usepackage{amsmath}
\usepackage{amsfonts}
\usepackage{amssymb}
\usepackage{amsbsy}
\usepackage{amsthm}

\usepackage[pdftex,colorlinks=true,urlcolor=blue,citecolor=black,anchorcolor=black,linkcolor=black]{hyperref}

\newtheoremstyle{wsc}
{3pt}
{3pt}
{}
{}
{\bf}
{}
{.5em}
{}

\theoremstyle{wsc}

\begin{document}

%
%

\pagestyle{fancyplain}

\thispagestyle{plain}
\firstPageHead{}

\chead{\fancyplain{}{\itshape Madad}}

\rhead{}
\cfoot{}
\renewcommand{\headrulewidth}{0pt} 

\makeatletter
\let\@internalcite\cite
\def\cite{\def\@citeseppen{-1000}%
    \def\@cite##1##2{(##1\if@tempswa , ##2\fi)}%
    \def\citeauthoryear##1##2##3{##1 ##3}\@internalcite}
\def\citeNP{\def\@citeseppen{-1000}%
    \def\@cite##1##2{##1\if@tempswa , ##2\fi}%
    \def\citeauthoryear##1##2##3{##1 ##3}\@internalcite}
\def\citeN{\def\@citeseppen{-1000}%
    \def\@cite##1##2{##1\if@tempswa, ##2)\else{}\fi}%
    \def\citeauthoryear##1##2##3{##1 (##3)}\@citedata}
\def\citeA{\def\@citeseppen{-1000}%
    \def\@cite##1##2{(##1\if@tempswa , ##2\fi)}%
    \def\citeauthoryear##1##2##3{##1}\@internalcite}
\def\citeANP{\def\@citeseppen{-1000}%
    \def\@cite##1##2{##1\if@tempswa , ##2\fi}%
    \def\citeauthoryear##1##2##3{##1}\@internalcite}
\def\shortcite{\def\@citeseppen{-1000}%
    \def\@cite##1##2{(##1\if@tempswa , ##2\fi)}%
    \def\citeauthoryear##1##2##3{##2 ##3}\@internalcite}
\def\shortciteNP{\def\@citeseppen{-1000}%
    \def\@cite##1##2{##1\if@tempswa , ##2\fi}%
    \def\citeauthoryear##1##2##3{##2 ##3}\@internalcite}
\def\shortciteN{\def\@citeseppen{-1000}%
    \def\@cite##1##2{##1\if@tempswa, ##2\else{}\fi}%
    \def\citeauthoryear##1##2##3{##2 (##3)}\@citedata}
\def\shortciteA{\def\@citeseppen{-1000}%
    \def\@cite##1##2{(##1\if@tempswa , ##2\fi)}%
    \def\citeauthoryear##1##2##3{##2}\@internalcite}
\def\shortciteANP{\def\@citeseppen{-1000}%
    \def\@cite##1##2{##1\if@tempswa , ##2\fi}%
    \def\citeauthoryear##1##2##3{##2}\@internalcite}
\def\citeyear{\def\@citeseppen{-1000}%
    \def\@cite##1##2{(##1\if@tempswa , ##2\fi)}%
    \def\citeauthoryear##1##2##3{##3}\@citedata}
\def\citeyearNP{\def\@citeseppen{-1000}%
    \def\@cite##1##2{##1\if@tempswa , ##2\fi}%
    \def\citeauthoryear##1##2##3{##3}\@citedata}
%
%
%
\def\@citedata{%
    \@ifnextchar [{\@tempswatrue\@citedatax}%
                  {\@tempswafalse\@citedatax[]}%
}

\def\@citedatax[#1]#2{%
\if@filesw\immediate\write\@auxout{\string\citation{#2}}\fi%
  \def\@citea{}\@cite{\@for\@citeb:=#2\do%
    {\@citea\def\@citea{, }\@ifundefined
       {b@\@citeb}{{\bf ?}%
       \@warning{Citation `\@citeb' on page \thepage \space undefined}}%
{\csname b@\@citeb\endcsname}}}{#1}}%

%
\def\@citex[#1]#2{%
\if@filesw\immediate\write\@auxout{\string\citation{#2}}\fi%
  \def\@citea{}\@cite{\@for\@citeb:=#2\do%
    {\@citea\def\@citea{; }\@ifundefined
       {b@\@citeb}{{\bf ?}%
       \@warning{Citation `\@citeb' on page \thepage \space undefined}}%
{\csname b@\@citeb\endcsname}}}{#1}}%

%
\def\@biblabel#1{}
\makeatother



\newdimen\bibindent
\bibindent=0.0em
\def\thebibliography#1{\section*{\refname}\list
   {}{\settowidth\labelwidth{[#1]}
   \leftmargin\parindent
   \itemindent -\parindent
   \listparindent \itemindent
   \itemsep 0pt
   \parsep 0pt}
   \def\newblock{}
   \sloppy
   \sfcode`\.=1000\relax}


\setlength{\baselineskip}{12.7pt}

\title{Efficient Distance Pruning for process suffix comparison in Prescriptive Process Monitoring}

\author{\begin{center}Sarra Madad\textsuperscript{1,2}\\
[11pt]
\textsuperscript{1}Université de Technologie de Troyes, LIST3N Research Unit, Troyes, FRANCE\\
\textsuperscript{2}QAD Process Intelligence, Courbevoie, FRANCE\end{center}
}

\maketitle

\vspace{-12pt}

\section*{ABSTRACT}
Prescriptive process monitoring seeks to recommend actions that improve process outcomes by analyzing possible continuations of ongoing cases. A key obstacle is the heavy computational cost of large-scale suffix comparisons, which grows rapidly with log size. We propose an efficient retrieval method exploiting the triangle inequality: distances to a set of optimized pivots define bounds that prune redundant comparisons. This substantially reduces runtime and is fully parallelizable. Crucially, pruning is exact: the retrieved suffixes are identical to those from exhaustive comparison, thereby preserving accuracy. These results show that metric-based pruning can accelerate suffix comparison and support scalable prescriptive systems.

\section{INTRODUCTION : PRESCRIPTIVE PROCESS MONITORING FOR SUFFIX PREDICTION}
Suffix prediction plays a central role in process analytics, as it enables estimating how an ongoing execution may unfold under different continuations. In the context of process mining, a suffix simply denotes the sequence of future activities of a case until its completion. Building on this capability, prescriptive approaches aim not only to anticipate future behavior but also to recommend interventions that improve process performance \shortcite{Weinzierl_2020}. In this setting, the system must evaluate how different possible continuations of the current execution would affect key performance indicators (KPIs), enabling the recommendation of a next-best action. A central mechanism to achieve this is the search for contrasting suffixes: process continuations from past cases that diverge in their outcomes. By comparing these suffixes, the system can infer which actions tend to lead to favorable trajectories. This strategy entails numerous pairwise distance computations, which become prohibitive as event logs grow \cite{berti2019increasingscalabilityprocessmining}. To mitigate this, we apply a triangle inequality–based pruning method \cite{JEROMIN198970}, where distances to a small set of pivots define bounds that allow discarding many redundant comparisons. This significantly reduces computation while preserving exactness.

\section{TRIANGULAR INEQUALITY ACCELERATION}
Let $\mathcal{S}$ be the countable set of process suffixes, $d:\mathcal{S}\times\mathcal{S}\to[0,\infty)$ a distance function (e.g., Euclidean or cosine distance), $\tau\in[0,\infty)$ a threshold, and $P=\{z_1,\dots,z_K\}\subset\mathcal{S}$ a finite pivot set chosen to cover $\mathcal{S}$ (e.g., minimizing $R(P)=\max_{x}\min_{z\in P} d(x,z)$). For any suffixes $x, y \in \mathcal{S}$ and a pivot $z \in \mathcal{S}$, the triangle inequality gives:
\[
\big|d(x,z)-d(y,z)\big| \;\le\; d(x,y) \;\le\; d(x,z)+d(y,z).
\]

By introducing a set of $K$ pivots $P=\{z_1,\dots,z_K\}$, $P = \{z_1, \dots, z_K\} \subset \mathcal{S}$, we can refine these bounds as follows:
\[
\max_{1 \leq k \leq K} \big|d(x,z_k)-d(y,z_k)\big| \;\le\; d(x,y) \;\le\; \min_{1 \leq k \leq K} \big(d(x,z_k)+d(y,z_k)\big).
\]

This allows pruning: if the lower bound already exceeds a search threshold $\tau$ (e.g., for $k$-nearest neighbors \cite{1053964}), then computing $d(x,y)$ explicitly is unnecessary. Similarly, if the upper bound is below $\tau$, the pair $(x,y)$ can be accepted directly.

The effectiveness of this approach depends on the choice of pivots. A common strategy is to select pivots that ``cover'' the space of suffixes. This can be formalized as a $k$-center problem:
\[
\min_{P \subset \mathcal{S},\, |P|=K} \;\; \max_{x \in \mathcal{S}} \min_{z \in P} d(x,z),
\]
which seeks pivots that minimize the maximum distance of any suffix to its closest pivot. Although NP-hard, this objective can be approximated efficiently using a greedy farthest-point heuristic, which iteratively selects the suffix farthest from the already chosen pivots. This ensures that the selected pivots are well spread across the dataset, tightening the bounds and increasing pruning efficiency.

\textbf{Example.} Suppose we have two suffixes $x, y$, a threshold $\tau=4$, and two pivots $z_1, z_2$ with precomputed distances:
\[
d(x,z_1)=2,\quad d(x,z_2)=6,\quad d(y,z_1)=7,\quad d(y,z_2)=9.
\]
Then the lower bound is
\[
\max\{|2-7|, |6-9|\} = \max\{5, 3\} = 5.
\]
Since $5 > \tau$, the comparison between $x$ and $y$ can be discarded without computing $d(x,y)$.

At scale, this method relies on precomputing a distance matrix of size $|\mathcal{S}| \times K$ (suffixes $\times$ pivots), which can be reused across queries. In practice, a small number of well-chosen pivots often suffices to prune a large fraction of candidate comparisons, yielding substantial computational savings while preserving exactness.

\section{EVALUATION}
Processing the full dataset of about 150,000 suffixes originally required nearly 89 hours. With the new approach, batches of 500 suffixes take about 2.5 hours and the method is fully parallelizable. By construction, pruning is lossless: retrieved suffixes always match the baseline of exhaustive pairwise comparison, a result confirmed empirically with 100\% accuracy.

\section{CONCLUSION}
This approach addressed the scalability challenge of suffix retrieval in prescriptive process monitoring. By leveraging the triangle inequality with optimized pivot selection and batching strategies, the proposed approach drastically reduces the number of distance computations while preserving accuracy. The method is fully parallelizable and therefore well-suited for large-scale event logs.

\section*{ACKNOWLEDGMENTS}
The author wishes to express sincere appreciation to her Ph.D. supervisors, Frédéric Bertrand (CNAM) and Myriam Maumy (EHESP), and to her industrial mentor, Yoann Valero, for their continuous advice and encouragement. The study was conducted as part of a CIFRE Ph.D. fellowship supported by QAD Inc.

\footnotesize
\bibliographystyle{wsc}

\bibliography{demobib}

\end{document}